\documentclass[%
 preprint, 
 prl
 amsmath,amssymb,
 aps,
superscriptaddress
]{revtex4-2}

\usepackage{graphicx}
\usepackage{dcolumn}
\usepackage{bm}
\usepackage{xcolor}
\usepackage{amsmath}
\usepackage{soul}

\newcommand{\VTNCE}{V[TCNE]$_x$}

\begin{document}

\title{Strong photon-magnon coupling using a lithographically defined organic ferrimagnet}

\author{Qin Xu}
\affiliation{Department of Physics, Cornell University, Ithaca NY 14853}
 
\author{Hil Fung Harry Cheung}
\affiliation{Department of Physics, Cornell University, Ithaca NY 14853}

\author{Donley S. Cormode}
\affiliation{Department of Physics, The Ohio State University, Columbus, OH 43210}

\author{Tharnier O. Puel}
\affiliation{Department of Physics and Astronomy, University of Iowa, Iowa City, IA 52242}

\author{Huma Yusuf}
\affiliation{Department of Physics, The Ohio State University, Columbus, OH 43210}

\author{Michael Chilcote}
\affiliation{School of Applied and Engineering Physics, Cornell University, Ithaca NY 14853}

\author{Michael E. Flatt\'e}
\affiliation{Department of Physics and Astronomy, University of Iowa, Iowa City, IA 52242}

\author{Ezekiel Johnston-Halperin}
\affiliation{Department of Physics, The Ohio State University, Columbus, OH 43210}

\author{Gregory D. Fuchs}
\affiliation{School of Applied and Engineering Physics, Cornell University, Ithaca NY 14853}

\date{\today}

\begin{abstract}
We demonstrate a hybrid quantum system composed of superconducting resonator photons and magnons hosted by the organic-based ferrimagnet vanadium tetracyanoethylene (\VTNCE). Our work is motivated by the challenge of scalably integrating an arbitrarily-shaped, low-damping magnetic system with planar superconducting circuits, thus enabling a host of quantum magnonic circuit designs that were previously inaccessible.  For example, by leveraging the inherent properties of magnons, one can enable nonreciprocal magnon-mediated quantum devices that use magnon propagation rather than electrical current. We take advantage of the properties of \VTNCE, which has  ultra-low intrinsic damping, can be  grown at low processing temperatures on arbitrary substrates, and can be patterned via electron beam lithography.  We demonstrate the scalable, lithographically integrated fabrication of hybrid quantum magnonic devices consisting of a thin-film superconducting resonator coupled to a low-damping, thin-film \VTNCE~microstructure.  Our devices operate in the strong coupling regime, with a cooperativity as high as 1181(44) at T$\sim$0.4 K, suitable for scalable quantum circuit integration. This work paves the way for the exploration of high-cooperativity hybrid magnonic quantum devices in which magnonic circuits can be designed and fabricated as easily as electrical wires. 

\end{abstract}

\maketitle

\newpage


Hybrid quantum systems are attractive for emerging quantum technologies because they take advantage of the distinct properties of the constituent excitations~\cite{Xiang2013,Clerk2020}.  This is important because no single quantum system is ideal for every task, e.g. scalable quantum information processing, quantum sensing, long-lived quantum memory, and long-range quantum communication all have different requirements.  Some hybrid systems that have been explored are microwave photons hybridized with spins~\cite{verdu2009strong, schuster2010high, kubo2010strong, kubo2011hybrid, amsuss2011cavity, chiorescu2010magnetic, bushev2011ultralow, abe2011electron, zhu2011coherent, viennot2015coherent,Tabuchi2015, Bienfait2016, xu2022quantum}, optical photons hybridized with atomic degrees of freedom~\cite{thompson2013coupling, spillane2005ultrahigh, goban2014atom, kimble1998strong}, and superconducting qubits hybridized with phonons~\cite{o2010quantum, chu2018creation, satzinger2018quantum, arrangoiz2019resolving}.  In creating hybrid systems, it is advantageous to operate in the strong-coupling, low-loss regime, where the relaxation rates of the two distinct quantum systems are exceeded by the coupling rate between them. This allows the hybrid system to operate as a quantum interconnect, wherein quantum information can be passed from one excitation to another~\cite{Clerk2020}. 
Thus, a central challenge is to couple distinct quantum systems strongly, with all elements maintaining long coherence times.  An equally critical  challenge is to fabricate the hybrid quantum devices using scalable and integrable approaches so that their engineered properties can be used in applications.

Combining magnons (quantized spin waves) and microwave photons into a hybrid quantum system harnesses the unique  properties of magnetic materials~\cite{Soykal2010,harder2021coherent,Yuan2022,Shim2020}.  For example, magnetic materials break time-reversal symmetry, providing a natural opportunity to create nonreciprocal  devices.  Circulators --- classical but nonreciprocal circuit components based on ferromagnetic resonance --- are indispensable for superconducting qubits. Magnon-cavity interactions provide new opportunities for nonreciprocal behavior~\cite{Wang2019,Zhang2020}, which is useful also for quantum devices~\cite{Lodahl2017,Bernier2017}. Additionally, magnetic materials are promising for coupling to long-lived quantum spin systems~\cite{candido2020predicted,solanki2022electric,lee2020nanoscale,mccullian2020broadband}, potentially enabling quantum interconnects between important quantum technologies based on microwave photons and on color center spins~\cite{Awschalom2021}. 

The challenge of integrating a microwave resonator with a low-damping magnetic material that can be lithographically patterned has limited the scalability of hybrid quantum magnonic systems.  Landmark initial demonstrations used large (millimeter-scale) crystals of the ferrimagnetic insulator yttrium iron garnet (YIG) because of its record-low damping, even at temperatures below 1~K for bulk crystals~\cite{Huebl2013,Tabuchi2014}.  These efforts used either copper three-dimensional microwave cavities~\cite{Tabuchi2014,Zhang2014,Haigh2015,Wang2016} or a planar waveguide cavity~\cite{Huebl2013}.  With focused ion beam milling and nanomanipulators, a micron-scale YIG crystal has been integrated with a superconducting resonator~\cite{Baity2021}.  Unfortunately, direct lithographic integration of thin-film YIG with superconducting planar circuits is an outstanding challenge, likely because lattice-matched substrates such as gadolinium gallium garnet are strongly paramagnetic and thus a lossy microwave substrate~\cite{maryvsko1989paramagnetic, maryvsko1991influence}. Another approach was the integration of permalloy (Ni$_{0.81}$Fe$_{0.19}$), a ferromagnetic alloy that can be directly grown and patterned on a planar cavity without the need for high-temperature processing or lattice matching, which substantially enhances the scalability of this hybrid system~\cite{Hou2019,Li2019}. Unfortunately, permalloy has orders-of-magnitude larger intrinsic magnetic damping as compared with high-quality bulk YIG crystals. 

In this work, we demonstrate an alternate pathway to a strongly-coupled hybrid magnonic system in which a low-damping magnetic film is patterned directly on a superconducting resonator under gentle growth conditions.  
We develop a process to integrate lithographically-patterned films of the organic-based ferrimagnet vanadium tetracyanoethylene (\VTNCE) with a planar superconducting microwave resonator. \VTNCE~has a typical Gilbert damping coefficient in the range $\alpha=(4-20)\times10^{-5}$~\cite{Franson2019,Cheung2021, bola2021fabrication}, which is comparable to YIG bulk crystals and high-quality YIG films grown on lattice matched substrates~\cite{lenk2011building, onbasli2014pulsed}. We demonstrate strong coupling between \VTNCE~magnons and resonator photons in two devices (3.6 GHz and 9.2 GHz) with a cooperativity as high as $\sim$10$^3$.  
This is critically enabling for integration and scaling, permitting future designs in which magnonic waveguides can be tailored as couplers or can mediate interactions between different quantum excitations. 
Focusing on the 3.6 GHz device, we present a detailed microwave transmission spectrum that reveals not only the expected avoided level crossing of the resonator mode and the uniform magnon mode (the Kittel mode, or simply magnon mode unless otherwise stated) but also the resonator mode hybridized with a discrete spectrum of excited magnon modes.
Finally, we study the relaxation rate of the hybrid system as a function of the system detuning in the frequency domain and the time domain.  This work represents a paradigmatic shift in hybrid magnonic quantum systems by establishing an integrated and scalable platform enabling arbitrary design of the magnonic elements.


The Hamiltonian describing a magnon mode coupled to a resonator mode is~\cite{Jaynes1963,Tavis1968,Tavis1969,Soykal2010,Tabuchi2014,Hou2019,Zhang2014}: 
\begin{equation}
    {\cal H}_0/\hbar = \omega_r \left(\hat{a}^\dagger \hat{a} + \frac{1}{2}\right) +\omega_{m}(B_0)\hat{b}^{\dagger}\hat{b}+g\left(\hat{b}^{\dagger}\hat{a}+\hat{b}\hat{a}^\dagger\right)
\label{H0}
\end{equation}
where the first two terms describe, respectively, the energy of resonator photons and magnons at a static field $B_0$, while the third term describes the photon-magnon 
coupling. We have used $\hat{a}$ and $\hat{a}^\dagger$ to describe the creation and annihilation of resonator photons, and $\hat{b}$ and $\hat{b}^\dagger$ to describe the creation and annihilation of  magnons --- these are derived from the Holstein-Primakoff transformation on spin operators~\cite{Aharoni-1996}. The collective coupling rate between the resonator photons and magnons is estimated as~\cite{Tabuchi2014} $g=g_s \sqrt{N}$, where $N$ is the number of spins within the magnetic material coupled to the resonator. One can estimate the single spin coupling rate from the geometry as~\cite{Hou2019, eichler2017electron}  $g_s = g_e \mu_B b_{rf} \omega_r / \sqrt{8 \hbar Z_r}$, where $b_{rf} = {\mu_0}/{2 w}$ is the magnitude of the magnetic field experienced by V[TCNE]$_x$ spins per unit current in an inductor wire of width $w$ when the spins are in close contact.  We have also used the electron Land\'e $g$ factor $g_e$, the Bohr magneton $\mu_B$, and the characteristic impedance of the resonator $Z_r=\sqrt{L/C}$. 
 
To design a hybrid system that is useful for quantum circuit integration, we desire a system operating in the strong-coupling, low-loss regime in which both the resonator damping rate $\kappa_r$ and the magnon damping rate $\kappa_m$ are smaller than $g$.  It is also useful to parameterize the system in terms of its cooperativity, $\mathcal{C} = {4g^2}/{\kappa_r \kappa_m}$, which exceeds 1 when the two systems are strongly coupled~\cite{Tabuchi2014}.  In that case, when we tune the system into resonance, the excitations are best described as hybrids of resonator photons and magnons. 

First we discuss the requirements of the magnetic material.  Low intrinsic (Gilbert) damping is critical for  minimizing $\kappa_m$.  Additionally, it is a major materials challenge to precisely pattern and integrate low-damping magnetic material with the superconducting resonator.  For this purpose \VTNCE~is advantageous because it can be grown on nearly any substrate without the need for high temperature processing~\cite{harberts2015chemical, zhu2016low, de2000cvd, pokhodnya2000thin}, giving it wide substrate compatibility.  Moreover, it can be patterned using electron beam lithography and lift-off techniques without compromising the ultra-low damping~\cite{Franson2019, zhu2020organic}, which overcomes the challenges of working with bulk-grown crystals and thus has considerable advantages for scaling and integration.  One of the unusual properties of \VTNCE~as a magnetic material is that it has a relatively low value of the saturation magnetization $\mu_0 M_s \sim$ 10~mT~\cite{trout2022probing, Franson2019, Yusuf2021}.  On one hand, this could be a disadvantage in  reaching a large $\sqrt{N}$ to enable strong coupling.  On the other hand, it is advantageous from a device design point of view because it allows one to work at comparatively small applied magnetic field, which avoids superconducting vortex formation.

For the microwave resonator we select a lumped-element design consisting of a planar interdigitated capacitor that is shorted by a narrow inductor wire.  This device is structurally similar to a transmon qubit that has a capacitor and a narrow inductor, except our structure does not include a Josephson junction as a part of the inductor.  Instead, our narrow inductor wire efficiently couples resonator excitations to the magnon mode when the magnetic material is patterned directly on the wire surface. The resonator mode concentrates the oscillating current through the wire, generating an Oersted magnetic field that excites the spins~\cite{Bienfait2016,Hou2019,Baity2021}.  Using our device's characteristic impedance $Z_r$=17.0(4.5)~$\Omega$ and inductor width $w$ = 10~$\mu$m, we estimate $g_s = 36(5)$~Hz.  Using $N = 2.195\times 10^{12}$ (from the magnetic volume described below, and $M_s$), the total coupling rate is estimated to be $g$ = 54(8) MHz.
 The resonator is capacitively coupled to a coplanar microwave feedline that we use to excite and detect the coupled resonator-magnon system. 
Microwave electromagnetic simulations of the bare resonator design are available in the supplementary materials.

The fabrication procedure must integrate the growth and patterning of inorganic superconducting films with organic-based \VTNCE~\cite{Fronig2015,Cheung2021}. We begin by sputtering 60~nm of Nb on a sapphire wafer and use photolithography and dry etching to form the microwave resonator. After wafer dicing, we spin a poly-methyl methacrylate (PMMA) bilayer on an individual die, which we expose using ebeam lithography.  Once developed and passivated, the resulting structure is a lift-off template for CVD-deposited \VTNCE~\cite{Franson2019}.  Inside an inert gas glovebox, we deposit the \VTNCE~at $\sim 60$~$^\circ$C~\cite{harberts2015chemical,yu2014ultra,zhu2016low} and then perform lift-off.  The \VTNCE~is then encapsulated by epoxy and a glass coverslip, stabilizing the material for weeks under ambient conditions~\cite{Fronig2015,Cheung2021}.  Figure 1(a) shows an optical micrograph of an encapsulated device with a resonator frequency $\omega_r/2\pi \sim$ 3.6~GHz.  A zoom into the narrow inductor shows the bright 10-$\mu$m-wide Nb wire with a dark gray strip of patterned \VTNCE~running down its center. The \VTNCE~strip is 300~nm thick, 6~$\mu$m wide and 600~$\mu$m long.

\begin{figure}
\includegraphics{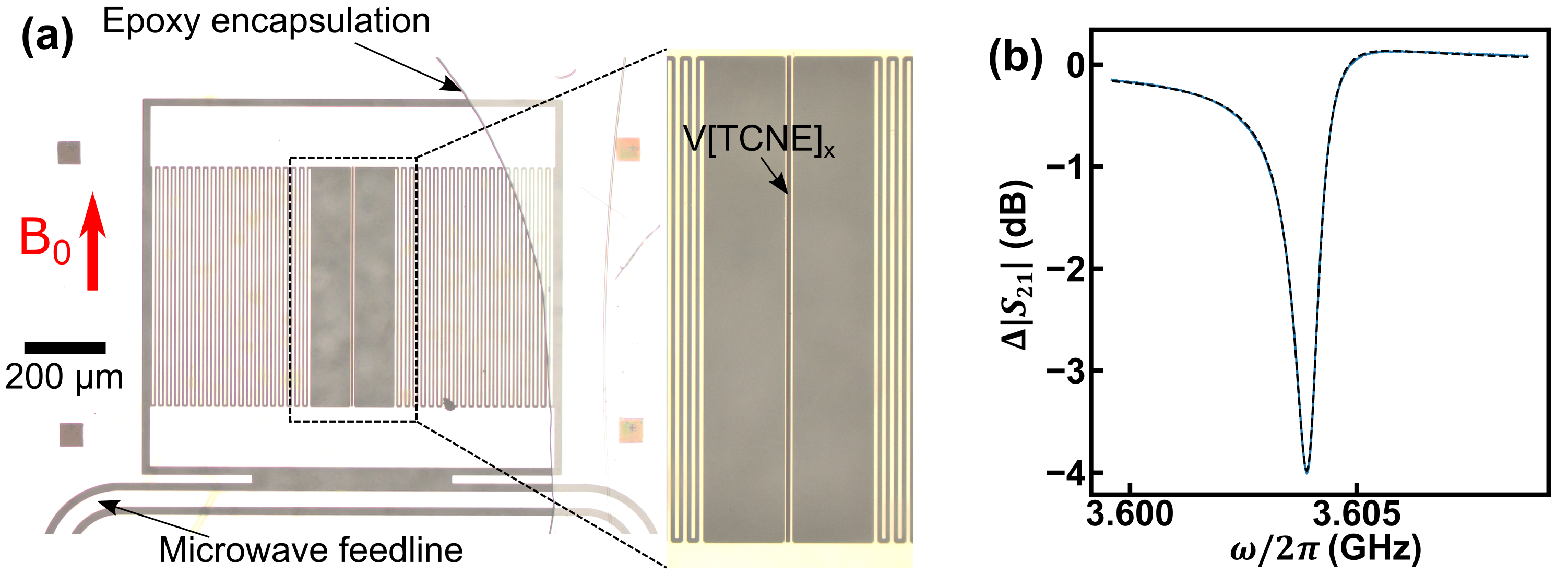}
\caption{(a) Microscope image of the 3.6~GHz resonator after \VTNCE~was deposited on its inductor wire. The \VTNCE~can be seen as a dark gray strip on the bright Nb wire at the center of the structure. The curved line is the border of the encapsulating epoxy. (b) Transmission spectrum (blue) and fit (black dashed curve) at 0.0809~T when the device is at 0.43~K.
}

\end{figure}


We start the electrical characterization of the device at 0.0809~T, a magnetic field at which the resonator and magnons are strongly detuned.  We measure the microwave transmission through the feedline using a vector network analyzer (VNA) at sample temperature $T$ = 0.43~K. The result is shown in Fig.~1(b). The transmission coefficient of a microwave resonator coupled to a transmission line can be modeled as~\cite{probst2015efficient}
\begin{equation}
    |S_{21}(\omega)| = \left|a \times \left(1 - \frac{(Q_l / |Q_c|) e^{i \phi}}{1 + 2iQ_l(\omega / \omega_{res} - 1)}\right)\right|, \label{S21}
\end{equation}
where $a$ is the attenuation coefficient, $\omega_{res} \approx \omega_r$ is the mode resonance frequency and $Q_l$ is the loaded quality-factor that is used to calculate the total damping rate via $\kappa_{l} = \omega_{res}/Q_l$. $|Q_c|$ is the magnitude of the coupling quality-factor that parameterizes the loss of photons from the resonator to the waveguide and $\phi$ is the phase of $Q_c$.

Fitting to Eqn.~\ref{S21} reveals $Q_l$ = 4302, $|Q_c|$ = 11200, and $\omega_r / 2\pi$ = 3.604~GHz.  Since these measurements are far detuned from the magnon resonance, the measured damping rate is approximately the resonator damping $\kappa_r = \omega_r/Q_l = 2\pi \times 0.8377$~MHz.  As discussed below, we find that $\omega_r$ and $\kappa_r$ have weak but non-zero magnetic field dependence.  We attribute this behavior to vortices in the superconducting film that can increase $\kappa_r$~\cite{santavicca2016microwave, vissers2015frequency}. 
We model these contributions phenomenologically, assuming that they vary linearly with magnetic field. As discussed below, we estimate $\kappa_r / 2\pi$ to be $0.902(32)$~MHz at the resonance field $B_{res}$ where $\omega_m(B_{res}) = \omega_r$. 

Next we tune the external magnetic field closer to the avoided level crossing between the microwave resonator mode and the magnon mode to study the coupling between the two systems. The coupled eigenfrequencies are~\cite{Huebl2013}
\begin{equation}
    \omega_\pm = \omega_r + \Delta/2 \pm \sqrt{\Delta^2 + 4g^2}/2,
    \label{omega}
\end{equation}
where $\Delta = \omega_m(B_0) - \omega_r$ is the system detuning, and the magnon frequency is $\omega_m(B_0)=\gamma \sqrt{B_0 (B_0 + \mu_0 M_{\text{eff}})}$ for the static magnetic field $B_0$  applied parallel to the long axis of the V[TCNE]$_x$ strip. Here $M_{\text{eff}} = M_s - H_k$ describes the difference between $M_s$ and the uniaxial anisotropy field $H_k$ of V[TCNE]$_x$. 
The transmission spectrum through the feedline will be modified by the resonator and its interactions with the magnons~\cite{harder2021coherent}. The total transmission due to line attenuation and magnon coupling is~\cite{Huebl2013}
\begin{equation}
    S_{21}(\omega, B_0) = S_{21,0}(\omega) 
    \left(1+\frac{(\kappa_{ext} / 2) e^{-i \phi}}{i(\omega - \omega_r)-\kappa_r / 2 +{g^2}\{{i[\omega - \omega_m(B_0)]-\kappa_m / 2}\}^{-1}}\right), \label{2dipsS21}
\end{equation}
where $\kappa_{ext}$ describes the coupling rate between the feedline and the resonator, and $S_{21,0}(\omega)$ is the background transmission.  

Figure 2(a) shows the corresponding experimental data $\Delta|S_{21}|=20$log$_{10}(|S_{21}|/|S_{21,0}|)$,
measured at $T$ = 0.43~K as a function of magnetic field and frequency. It shows a strong avoided level crossing --- direct evidence of strong coupling between resonator photons and magnons. These data are acquired with a VNA using $-$75~dBm of microwave power at the sample, which is far below the resonator's nonlinear power level, corresponding to about $3.9 \times 10^6$ resonator photons for large $\Delta$~\cite{schneider2021quantum} (see supplementary materials for details). At each field, we extract the resonance frequencies $\omega_\pm$ and resonance linewidths $\kappa_\pm$ of the upper and lower branch by fitting to Eqn.~\ref{S21}.

\begin{figure}
\includegraphics{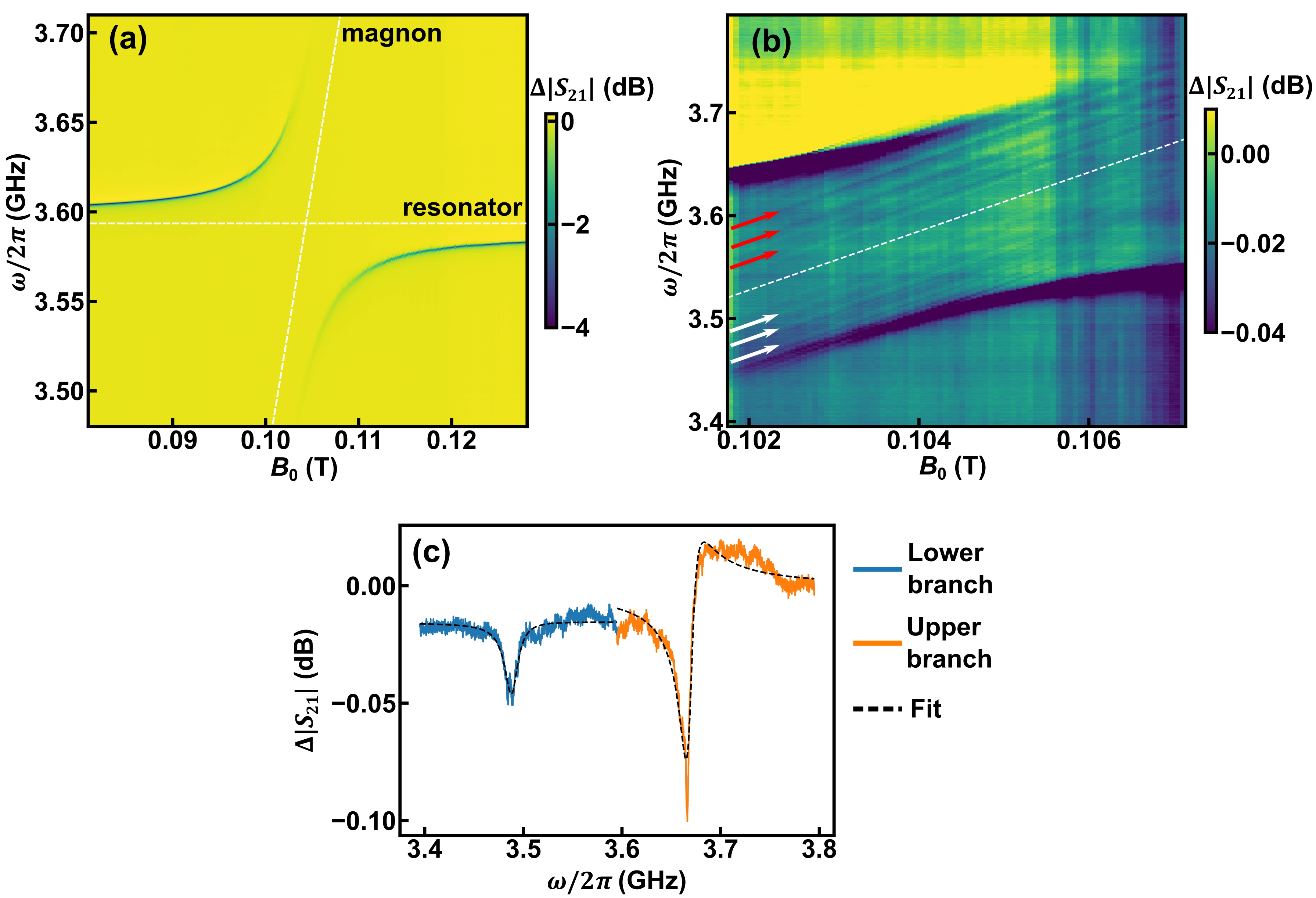}
\caption{Microwave measurement of \VTNCE~coupled to a 3.6~GHz resonator at 0.43~K. (a) $\Delta |S_{21}|$ plotted as a function of magnetic field and frequency. The white dashed lines mark the resonator and the magnon mode frequencies in the case where they are not coupled. (b) $\Delta |S_{21}|$ acquired in finer field steps near resonance field. We attribute the faint diagonal lines with slope 28~GHz/T to be $k\neq 0$ magnon modes. (c) $\Delta |S_{21}|$ line cut at $B_0 = B_{res}$ (e.g. $\Delta = 0$) showing the two branch resonances at $\omega_+$ and $\omega_-$. The dashed lines are fits as discussed in the text.
}
\label{fig2}
\end{figure}

To resolve the features at small $\Delta$ and to determine $g$ and $\kappa_m$ accurately, we measure $S_{21}$ with finer magnetic field steps. The data are presented in Fig.~2(b). Qualitatively, the strongest features are the two hybrid photon-magnon branches dispersing as $\omega_\pm$ given by Eqn. \ref{omega}. Additionally, we observe fainter modes that are linearly dispersing with a slope of 28~GHz/T. In contrast to the uniform magnon mode, we attribute those to magnon modes with a finite wavevector $k$~\cite{Stenning:13, BhoiB2014Sopc, serga2010yig}, which we discuss in further detail below.  

Figure 2(c) shows a line cut of $\Delta |S_{21}|$ as a function of frequency acquired at $B_0 = B_{res}$ (e.g. $\Delta = 0$), with corresponding fits to Eqn.~\ref{S21} for each branch. 
The two resonances are the $\omega_\pm$ as defined by Eqn.~\ref{omega}. We note that although $\omega_{-}$ has a symmetric resonance lineshape, $\omega_{+}$ has a Fano-like lineshape.  Nevertheless, we can use the splitting between these resonances to find $g=[\omega_+(B_{res})-\omega_-(B_{res})]/2 = 2\pi \times [90.31(8)]$~MHz.  The damping rate of the two branches, $\kappa_\pm$ are related to resonator and magnon damping rates via~\cite{harder2021coherent}
\begin{equation}
\kappa_\pm = [\kappa_r/2 + \kappa_m/2 \mp \text{Im} \sqrt{(-\omega_r+i\kappa_r/2+\omega_m- i\kappa_m/2)^2 + 4g^2}].
\end{equation}
From the linewidths in Fig. 2(c) we find $\kappa_-(B_{res})/2\pi  = 16.37(29)$~MHz and $\kappa_+(B_{res})/2\pi = 15.15(17)$~MHz. 
In the strong coupling regime ($g\gg \kappa_r,\kappa_m$), $\kappa_+=\kappa_- =(\kappa_r+\kappa_m)/2$. Therefore we take the average of experimental $\kappa_+,\kappa_-$ to estimate $(\kappa_r+\kappa_m)/2$. Together with $\kappa_r(B_{res})/2\pi$ = 0.902(32)~MHz, we estimate $\kappa_m/2\pi = 30.62(34)$~MHz.
We find that this device operates with $\mathcal{C} = 1181(44)$ and fulfills $g > \kappa_r, \kappa_m$.

\begin{figure}
\includegraphics{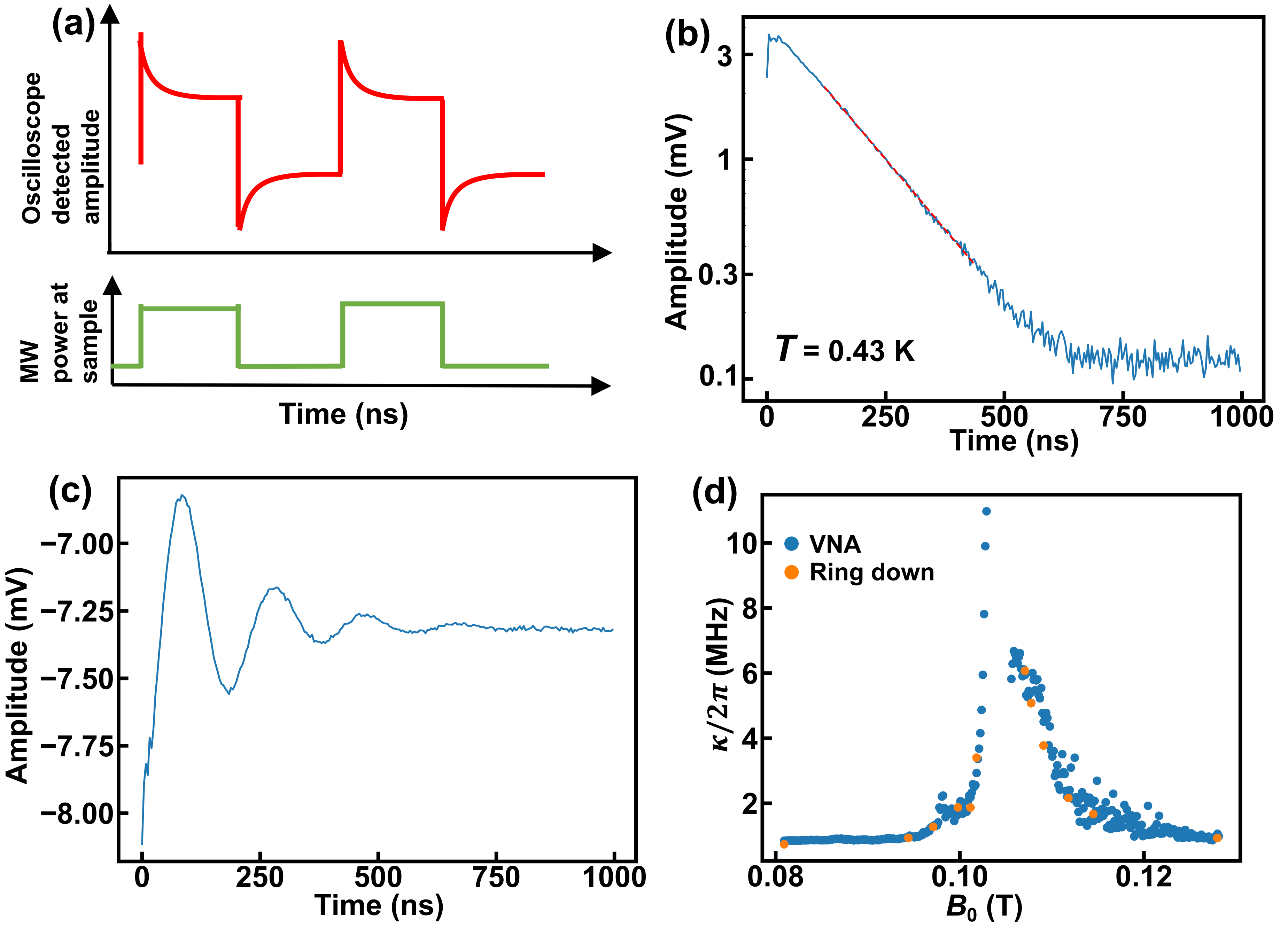}
\caption{Ring-down experiment with homodyne circuit for the 3.6~GHz sample at 0.43~K, $-$65~dBm excitation power. (a) Schematic plot of how the homodyne detected signal (red) changes with respect to the microwave excitation power at the sample (green). When the microwave excitation power turns on (off), system will ring up (down) to steady-state.  (b) Amplitude vs time plot of ring-down at 0.101~T for on resonance excitation. The fitted voltage ring-down time is 170.0(2.6)~ns. (c) Amplitude vs time plot with 5~MHz detuning. (d) Total system damping calculated from ring-down experiments (orange) and VNA experiments (blue).
}

\end{figure}

Having probed the system in the frequency domain, we now turn to measurements of system relaxation in the time domain.  For this, we apply pulsed microwave excitation to the feedline and detect the ring-up/-down response in the time domain using a homodyne circuit.  The full circuit and measurement protocol is discussed in the supplementary materials. The microwave pulses are long enough to excite the system into driven steady-state oscillations. When the microwave drive turns off suddenly, the system will continue to oscillate, however, it will do so at its natural frequency.  Its relaxation will include both intrinsic relaxation into the environment and relaxation through radiation into the feedline that we detect.  We amplify and mix this signal with the reference tone, and digitize the result. Figure 3(a) shows the schematic plot of microwave power at the sample and the oscilloscope detected voltage vs time.

Figure 3(b) shows a log-linear plot of amplitude relaxation after driving the system with $-$65~dBm of microwave power at the resonance frequency $\omega_+$ at $T$ = 0.43~K and $B_0 = 0.101$~T.  Under these conditions, the hybrid state has more resonator character than magnon character, however, $\omega_+ - \omega_r = 2\pi \times$ 43.9~MHz $\sim g/2$, indicating substantial magnon participation. We observe exponential relaxation with a voltage ring-down time constant of 170.0(2.6)~ns. Therefore, the energy relaxation time constant is $\tau = 85.0(1.3)$~ns, corresponding to a decay rate of $\kappa_+ / 2\pi= 1.872(29)~$MHz. We can also detune the microwave excitation with respect to the hybrid mode. In Fig.~3(c) we plot ring-down data acquired at the same field, except using a driving frequency that is detuned by 5~MHz from $\omega_+ /2\pi$. In this case, the homodyne signal beats with respect to the reference oscillator, giving rise to a decaying sinusoidal response.  We recover a decay rate of 1.924(28)~MHz on top of the 5~MHz beat, which is consistent with the on-resonance driving result.

To explore the relaxation rates at different detunings $\Delta$, we perform on-resonance ring-down measurements as a function of $B_0$.  We plot the resulting damping rates along with the damping rates extracted from VNA full-width at half-maximum (FWHM) linewidth measurements vs $B_0$ in Fig.~3(d). We see excellent agreement between the two methods, indicating that our VNA measurements are acquired in the linear response regime. Figure~3(d) underscores that the magnon damping rate is one order of magnitude larger than the bare resonator damping rate. The damping rate of the upper (lower) branch is increasing when the field is tuned closer to $B_{res}$ from below (above), since the hybrid state has a larger participation from the magnon mode as $\Delta$ approaches 0. The upper branch damping rate at 0.0809~T and the lower branch damping rate at 0.1281~T are assumed to be the resonator damping rates with no magnon contribution since these fields are far away from $B_{res}$. By linear interpolation, we estimate $\kappa_r / 2\pi$ to be $0.902(32)$~MHz at $B_{res}$, which we used above to calculate $\mathcal{C}$.  


We now turn our attention to the additional magnon modes that are evident in Fig. \ref{fig2}(b), i.e., the lines indicated by arrows within the avoided level crossing region.  For reference, we plot a dashed line showing the uniform magnon mode position if it were not coupled to the resonator. It separates the additional modes into higher frequency (red arrows) and lower (white arrows) frequency modes, which likely have distinct origins.
Here, we treat all magnon modes as spinwave modes characterized by a wavevector $\boldsymbol{k}$ and quantized by the boundary conditions at the surfaces of the \VTNCE. The case $k=0$ refers to the uniform magnon mode, while the additional modes are $k\neq 0$ magnon modes.
The width and length of the \VTNCE~strip are long enough that quantization constrained in those directions cannot be resolved; only modes that are quantized in the thickness direction ($|\boldsymbol{k}| \equiv k$ perpendicular to the magnetic field) are resolvable. 
We assign the features at frequencies above the dashed line to be thickness quantized $k \ne 0$ magnon modes. We do not have a definitive assignment for the modes lying at frequencies below the dashed line.  A more complete discussion can be found in the supplemental material.

\begin{figure}
\hfill{}\includegraphics[width=0.9\textwidth]{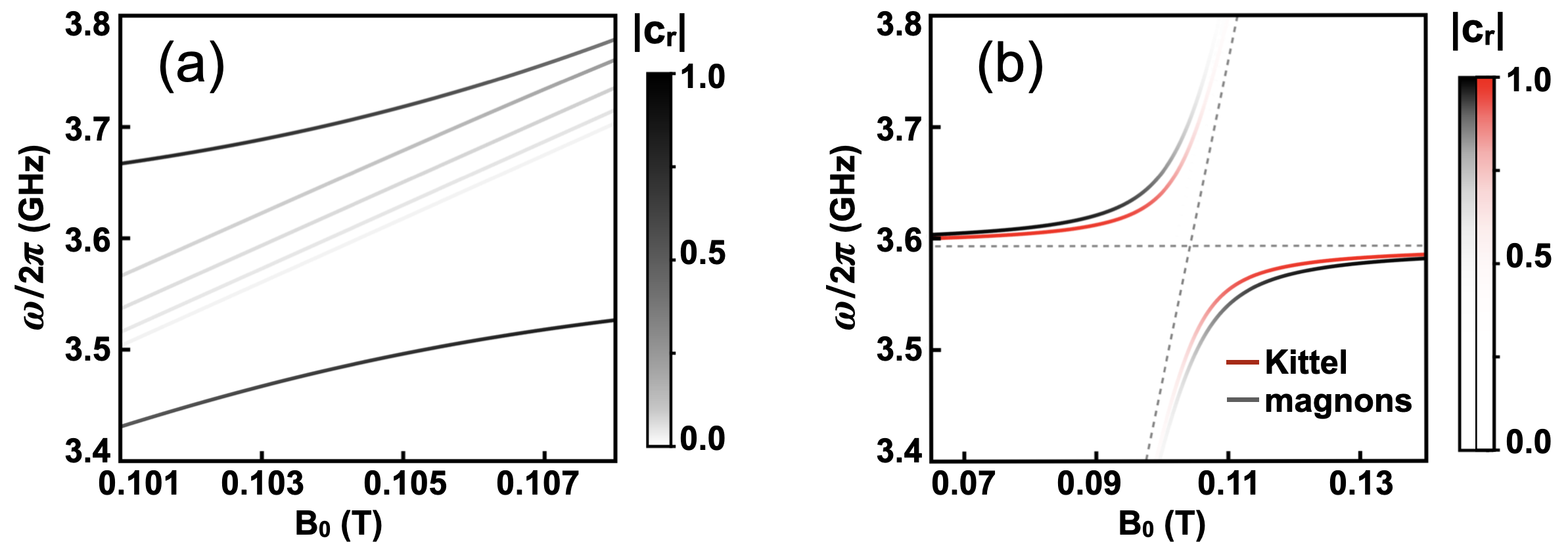}\hfill{}
\caption{(a) Energy spectrum of ${\cal H}\left|\psi_{i}\right\rangle =\hbar\omega_{i}\left|\psi_{i}\right\rangle $ colored by the coefficient $\left|\psi_{i}\right\rangle =\left(c_{r}^{(i)},\ldots\right)$ related to the resonator. This simulation used realistic values as follows: $L=300$~nm, $\gamma/2\pi=28$~GHz/T, $\mu_{0}M_{\text{eff}}=53.7$~mT, $\lambda_{\text{ex}}^{2}=0.25\times10^{-16}~\text{m\ensuremath{^{2}}}$, $\omega_{r}/2\pi=3.593$~GHz, $g/2\pi=90$~MHz, and $g_{n}=g/(n+1)$ for $n=1,\ldots,4$. (b) The results labeled by `magnons' repeat the plot (a), while the red lines include only the coupling between the uniform magnon mode (Kittel mode) and the resonator mode (both independently identified by the dashed lines).}
\label{fig: linear Hamiltonian}
\end{figure}

To better understand the transmission spectrum, we now discuss a model for multiple magnon modes coupled to the resonator. We extend the Hamiltonian ${\cal H}_0$ given in Eqn.~\ref{H0} by adding a term ${\cal H}_{s}$ that describes the $k\neq 0$ magnon modes as
\begin{align}
{\cal H}_{s}/\hbar & =\sum_{n=1}\omega_{n}\hat{s}_{n}^{\dagger}\hat{s}_{n}+\left(\sum_{n=1}g_{n}(\hat{s}_{n}^{\dagger}\hat{a}+\hat{s}_n \hat{a}^\dagger)\right).
\end{align}
We introduce creation and annihilation operators $\hat{s}_n$ and $\hat{s}_n^\dagger$ for the $k \ne 0$ magnons, and their 
direct coupling to the resonator has strength $g_n$.

We consider magnon modes described by dipole-exchange interactions~\cite{Gurevich-Melkov-1996};  their frequency dispersion is given in Ref.~\cite{Kalinikos_1986},
\begin{equation}
\omega_{n}= \gamma \sqrt{\left(B_0 + \mu_{0}M_{\text{eff}} \lambda_{\text{ex}}^2 k_{n}^{2}\right)\left(B_0 +\mu_{0}M_{\text{eff}}+\mu_{0}M_{\text{eff}} \lambda_{\text{ex}}^2 k_{n}^{2}\right)}.
\end{equation}
The quantization index $n=1,2,\ldots$ is along the thickness direction, where the wavevectors are constrained by $k_{n}L=n\pi$. Setting $n=0$ recovers the frequency of the uniform magnon mode (or Kittel mode), while the extra terms are due to the presence of exchange interactions with amplitude described by the exchange-length constant $\lambda_{\text{ex}}$. The expression above assumes $B_{0}$ is oriented in the plane of $\text{V[TCNE]}_{x}$.

Magnon modes with $k\neq 0$ can be directly excited by the magnetic field generated by the inductor only if their spatially-averaged amplitude does not vanish; therefore, their coupling to the resonator is highly dependent on the spin-pinning boundary conditions~\cite{Franson2019, puszkarski2016surface}. For instance, there is no direct coupling for totally unpinned boundaries, while for complete pinning only even $n$-index modes couple. Here, we consider an intermediate situation where all $g_{n}$ are allowed to exist, which can happen for partial pinning \cite{PhysRevLett.122.247202}. To demonstrate the essential features of this interaction, we chose $g_{n}=g/(n+1)$ as a qualitative description of the features observed, which is likely an overestimate. We solve the equation ${\cal H}\left|\psi_{i}\right\rangle =\hbar\omega_{i}\left|\psi_{i}\right\rangle $ as a function of $B_{0}$ and plot the results in Fig. \ref{fig: linear Hamiltonian}(a). The two branches $\omega_{\pm}$ represent the hybridized modes due to the strong uniform magnon-resonator coupling. The $k \ne 0$ modes have weak $\Delta|S_{21}|$ (Fig.~2) because the $|\psi_i\rangle$ have small resonator amplitudes, however, they become stronger as their frequencies approach the $\omega_{+}$ branch (Fig.~\ref{fig: linear Hamiltonian}(a)). We truncate at $n=4$ capturing only spin states that lie within the avoided-crossing gap, however, we note that the spacing between the $k \ne 0$ magnon modes is  sensitive to the parameter $\lambda_{\text{ex}}$ through a linear dependence on $\lambda_{\text{ex}}^2$. The coupling of the $k\neq 0$ magnons with the resonator increases the total  gap between the two branches 
$\omega_{\pm}$, as shown in Fig.~\ref{fig: linear Hamiltonian}(b).

 These advances in our understanding of the quantum magnonic properties of \VTNCE\\highlight its potential for applications in quantum information devices and the substantial potential for further optimization.   In this study, the uniform mode magnon damping rate $\kappa_m$ at $T = 0.43$~K corresponds to a FWHM ferromagnetic resonance (FMR) linewidth of 1.09~mT, which is comparable to the room temperature 0.57~mT  linewidth of the ``witness" sample that was deposited at the same time as the \VTNCE~used in this device and measured at $\sim$9.8~GHz.  This points to the opportunity for improvement through \VTNCE~growth and patterning considering that room temperature thin film \VTNCE~FMR linewidths as narrow as 0.094~mT \cite{trout2022probing}, and ebeam patterned \VTNCE\\FMR linewidths as narrow as 0.12~mT, at 9.8~GHz have been demonstrated~\cite{Franson2019}. Moreover, prior measurements of the FMR linewidth at temperatures down to 5 K revealed a strong temperature-dependent strain effect, highlighting one avenue for improvement~\cite{Yusuf2021}. \VTNCE~is a promising material for quantum magnonic applications with properties that rival and exceed that of YIG, is compatible with a broad range of materials, and can be lithographically integrated with planar superconducting circuits.


In conclusion, we have demonstrated a hybrid quantum system composed of superconducting resonator photons and magnons in the strong coupling regime with $\mathcal{C} = 1181$.  A key advance of this work is the integration of lithographically patterned and low-damping magnetic material with a superconducting circuit platform, enabling new quantum technologies in which the electrical and the magnonic structures are designed and fabricated on an equal basis. This capability can lead to nonreciprocal quantum circuit elements, new forms of hybrid-system couplers, new opportunities for tunability, and the creation of transmon qubits with integrated magnonic properties.  Beyond quantum technologies, this hybrid system can offer sensitive readout of coherent magnonic excitations.  The broad design space offered by lithographic patterning allows experiments that selectively probe different magnon wavevectors and -- in combination with superconducting qubits -- enables the creation, manipulation, and detection of single magnons in a scalable, integrated platform. 


\section*{Acknowledgements}
We thank Brendan McCullian for useful conversations.  The resonator design and fabrication, \VTNCE~growth and growth optimization, and theory of uniform magnon mode coupling were supported through the Center for Molecular Quantum Transduction (CMQT), an Energy Frontier Research Center supported by the Department of Energy Office of Science, Basic Energy Sciences (DE-SC0021314).  The development and design of resonator-\VTNCE~integration and lithography, measurement techniques, and theoretical analysis of $k\neq 0$ magnon modes were supported by the Department of Energy Office of Science, Basic Energy Sciences Quantum Information Sciences program (DE-SC0019250).  All measurements were done using the CMQT low temperature facility at Cornell.  This work also made use of facilities at the Cornell NanoScale Facility, an NNCI member supported by the NSF (NNCI-2025233) and the Cornell Center for Materials Research Shared Facilities which are supported through the NSF MRSEC program (DMR-1719875). We acknowledge the support of the NanoSystems Laboratory User Facility which is supported by the Center for Emergent Materials, an NSF MRSEC (DMR-2011876).

\newpage
\bibliography{FuchsBib}

\newpage
\section*{Supplementary Materials}

\section{Device Simulations}

We simulate the resonator with Keysight PathWave Advanced Design System (ADS) software. Fig S1(a) shows the simulated transmission coefficient $\Delta|S_{21}|$ vs frequency for the 3.6~GHz resonator. The simulated resonance frequency is 3.933 GHz. We then simulate the device being driven at the resonance frequency and plot the time averaged magnitude of current density in the superconducting film. The result is shown in fig S1(b), in which red indicates larger current density and blue indicates smaller current density. This resonance mode has a large current density in the inductor wire for efficient coupling to magnon modes of a magnetic material deposited on the wire. After fabrication, we measure the resonator's transmission spectrum using a VNA at 0 field before V[TCNE]$_x$ deposition. The result is shown in fig S1(c), where the resonance is at 3.804~GHz with Q-factor ($Q_l$) of 4922. This resonance frequency is lower than the simulation because of the finite kinetic inductance of Nb~\cite{santavicca2016microwave, vissers2015frequency}, which is not included in the simulation. This kinetic inductance adds to the geometric inductance of the LC resonator and decreases the resonance frequency. Fig S1(d) shows the transmission spectrum at 0 field after V[TCNE]$_x$ deposition and encapsulation. The encapsulation epoxy and cover glass increase the effective dielectric constant of the resonator environment, which increases the capacitance $C$, resulting in a lower resonance frequency. We speculate that the decrease of Q (2546) with epoxy encapsulation is caused by the loss tangent of the epoxy and glass, and loss from the non-uniform \VTNCE~magnetization at 0 field. Then we increase the field to 0.0944~T to saturate \VTNCE~magnetization. Fig 1(b) in the main text shows the resulting transmission spectrum. We attribute the increase in Q (4302) to the increase of the uniformity of the \VTNCE~magnetization. 

\renewcommand{\thefigure}{S1}
\begin{figure}
\includegraphics{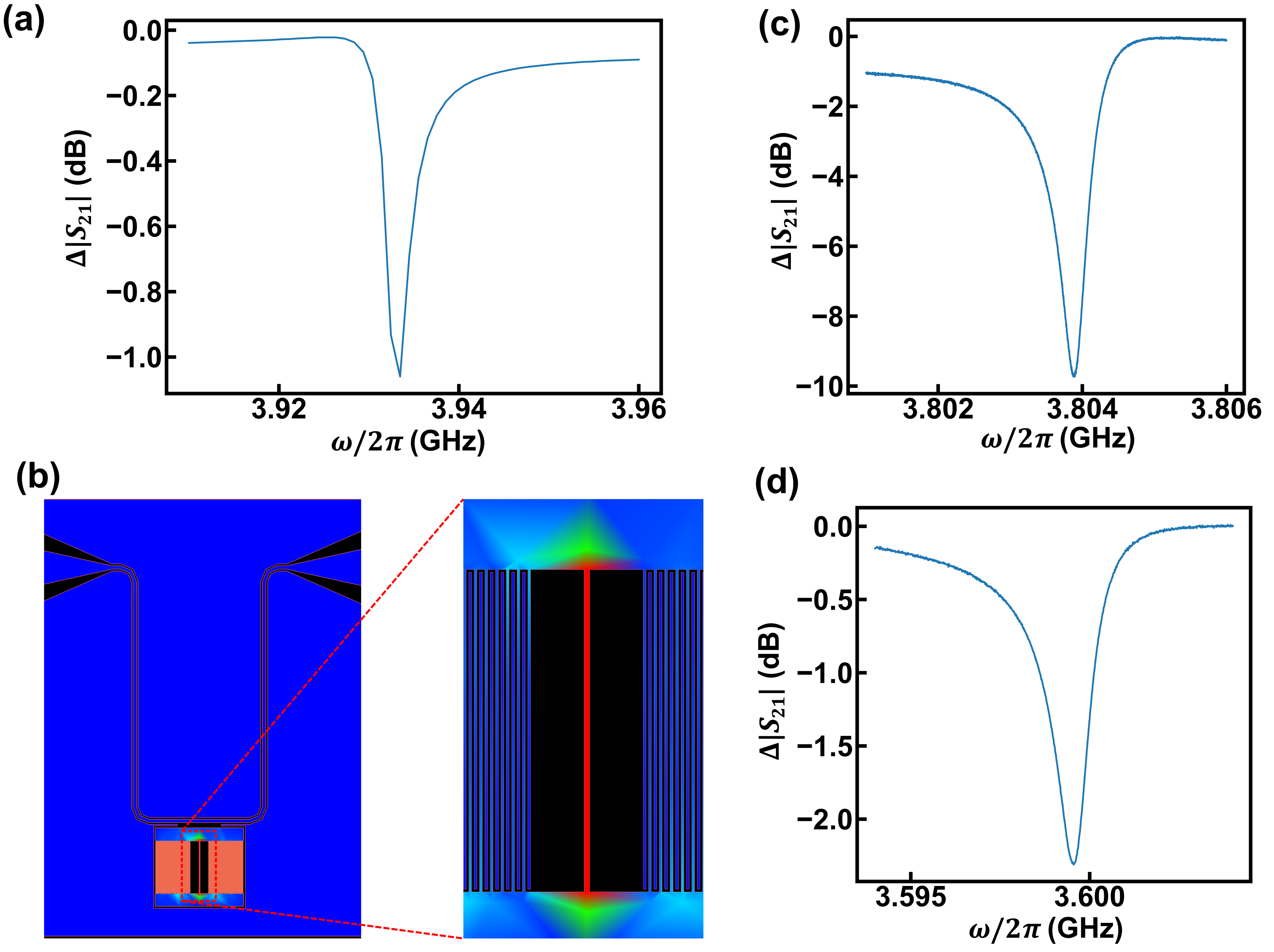}
\caption{(a) Microwave transmission $\Delta|S_{21}|$ from the ADS simulation of the 3.6~GHz resonator. We find a resonance frequency of 3.93~GHz. (b) The time averaged magnitude of current density plot of the resonator at the resonance frequency shown in (a). The zoom shows a large current density in the inductor wire. (c) Experimental measurement of $\Delta|S_{21}|$ at $B_0 = 0$ for a resonator at 0.43~K without \VTNCE. Fitting with Eqn. \ref{S21} gives $\omega_r / 2\pi = 3.804$~GHz and $Q_l = 4922$. The resonance frequency is lower than the simulation because of the finite kinetic inductance of Nb in the real device, which is not included in the simulation. (d) Experimental measurement of $\Delta|S_{21}|$ at $B_0 = 0$ for the \VTNCE-resonator device discussed in the main text, at 0.43~K. Fitting with Eqn. \ref{S21} gives $\omega_r / 2\pi = 3.600$~GHz and $Q_l = 2546$. 
}

\end{figure}

\section{Device Fabrication}

We deposit Nb as the superconducting material for our device on a sapphire wafer by sputtering at 600 Celsius. The resulting Nb is 60~nm thick and has a critical temperature (T$_c$) of 7.3~K. The high T$_c$ of Nb and the low loss tangent of sapphire help limit the damping rate of the superconducting device. 

We pattern the Nb film using photolithography followed by dry etching. We first spin-coat the wafer with photoresist, expose using a 5x stepper, and then develop. Then we etch the Nb using ion milling before stripping the photoresist.  

We use an electron beam lithography lift-off process to pattern the \VTNCE~as discussed in the main text. First we spin-coat individual resonator chips with  a bilayer of PMMA resist and expose the resist layer. We then develop the PMMA layers, leaving an opening for \VTNCE~deposition directly on the Nb. The \VTNCE~is deposited using CVD as discussed in the main text. The PMMA is then dissolved using dichloromethane, which leaves the \VTNCE~in the patterned regions untouched. The final step is to  encapsulate the resonator and patterned \VTNCE~using epoxy and a coverslip. The final dimensions of the \VTNCE~is 600~$\mu$m $\times$ 6~$\mu$m $\times$ 300~nm. 

\section{Resonator and Magnon Mode Saturation Power}

Using power-dependent $S_{21}$ measurements at $B_0 = 0.0994$~T, we determine the saturation power (of the kinetic inductance non-linearity) for the 3.6~GHz resonator to be larger than $-$65~dBm applied to the feedline port. To make sure we are not saturating the magnetic resonance, we also make power-dependent $S_{21}$ measurements near $B_{res}$. No power dependence is observed in the range of $-$65~dBm to $-$85~dBm. The data in Fig.~2 are acquired with $-$75~dBm of microwave power at the sample, which is well below saturation for both the resonator and the magnet. 

We now theoretically estimate the magnon precession cone angle. When we drive on resonance with the upper or lower branch $\omega_{res}$, the average number of excitations in the resonator-magnet system is~\cite{schneider2021quantum}

\begin{align*}
    \langle n \rangle = \frac{4P_{in}}{\hbar\omega_{res}^2} \frac{Q_l^2}{|Q_c|},
\end{align*}
where $P_{in} = -75$~dBm $= 3.16 \times 10^{-11}$~W is the excitation power at the sample, $Q_l$ is the loaded Q-factor and $Q_c$ is the coupling Q-factor. For the upper branch at $B_0 = B_{res}$, we find $Q_l = 242.1$, $|Q_c| = 23000$ and the resonance frequency $\omega_{res} / 2\pi = 3.669$~GHz from fitting to Eqn. \ref{S21}. Under these conditions, we estimate $\langle n \rangle = 5800$ and the average number of magnons $\langle n_{m} \rangle = \langle n \rangle / 2 = 2900$ at $B_0 = B_{res}$. 

The cone angle $ \theta $ of the uniform magnon mode satisfies
\begin{align*}
    1-\cos{\theta} \approx \frac{1}{2} \theta^2 = \frac{\langle n_{m} \rangle \hbar \omega_{res}}{\frac{1}{2}N \hbar \omega_{res}},
\end{align*}
where $N \approx 2.195 \times 10^{12}$ is the estimated number of \VTNCE~spins in the sample. 
Therefore, we estimate $\theta \approx 2 \sqrt{\langle n_{m} \rangle / N} = 7.2 \times 10^{-5}$~rad $= 0.0041^{\circ}$.

Similarly, in Fig. 1(b) with $\omega_{res}/2\pi = 3.604$~GHz, $Q_l = 4302$ and $|Q_c| = 11200$, we get the average number of resonator photons $\langle n_{r} \rangle \simeq \langle n \rangle = 3.9 \times 10^6$.

\section{Background Transmission}

The background transmission $S_{21,0}(\omega)$ is assumed to be independent of $B_0$ and is measured at some ($B_0$'s, $\omega$'s) that are far away from the upper or lower branch (see Fig. 2(a), 2(b)). In Fig. 2(a) for the 3.6~GHz device, $S_{21,0}(\omega)$ is measured from 3.48~GHz to 3.595~GHz at 0.1024~T, and from 3.595~GHz to 3.71~GHz at 0.1064~T. In Fig. 2(b), $S_{21,0}(\omega)$ is measured from 3.395~GHz to 3.495~GHz and from 3.595~GHz to 3.795~GHz at 0.1071~T, and 3.495~GHz to 3.595~GHz at 0.1017~T.

Similarly, in Fig. 5(b) for the 9.2~GHz device, $S_{21,0}(\omega)$ is measured from 8.9~GHz to 9.25~GHz at 0.2913~T, and from 9.25~GHz to 9.6~GHz at 0.3021~T. In Fig. 5(d), $S_{21,0}(\omega)$ is measured from 8.63~GHz to 9.23~GHz and from 9.47~GHz to 9.83~GHz at 0.2886~T, and 9.23~GHz to 9.47~GHz at 0.3021~T.

\section{Error Analysis}


Uncertainty is calculated via standard error analysis from least-squared fitting. We use the python package ``scipy.optimize.curve\_fit"~\cite{2020SciPy-NMeth}. The output includes the optimized values of all the fitting parameters and a 2-D array ``pcov" which gives the covarience matrix. We report the square root of the diagonal entries of the covariance matrix as the standard error.

\section{Extracted Magnetic parameters}
Using the data shown in Fig.~2(a), we fit $\omega_\pm$ by fitting to Eqn.~\ref{S21}. These results are then fit to Eqn.~\ref{omega} where $\omega_m$ is given by the Kittel formula with $\gamma / 2\pi = 28$~GHz/T. Also, we assume $\omega_r = \omega_{r0} + \gamma_r B_0$ and treat $\omega_{r0}$ and $\gamma_r$ together with $M_{\text{eff}}$ and $g$ as free parameters.  We have used $M_{\text{eff}} = M_s - H_k$, where $H_k$ is the uniaxial anisotropy field.  We obtain $\mu_0 M_{\text{eff}}$ = 53.614(63)~mT. Such a large $M_{\text{eff}}$ is likely caused by the large strain applied to the \VTNCE~due to differential thermal expansion~\cite{Yusuf2021}, which can induce a value of $H_k > M_s$.  

Next we determine the value of $B_{res}$ using the data shown in Fig. 2(b), again extracting $\omega_\pm$ by fitting to Eqn.~\ref{S21}. Figure S2 shows the extracted splitting $\omega_+ - \omega_-$ as a function of $B_0$. 
From Eqn.~\ref{omega} we know that $\omega_+ - \omega_- = \sqrt{\Delta^2 + 4g^2}$, where $\Delta = \omega_m - \omega_r = \gamma_{rm} \times (B_0 - B_{res})$, and $\gamma_{rm}$ is the change of $\omega_m - \omega_r$ per unit increase of $B_0$. Treating $g$, $B_{res}$ and $\gamma_{rm}$ as free parameters, we get $g / 2\pi = 90.43(8)~$MHz, $B_{res} = 0.103429(18)~$T and $\gamma_{rm} / 2\pi = 52.1(2.6)$~GHz/T. The best-fit curve is shown as the black dashed line in Fig.~S2. First, we find a value of $g$  that is consistent with the value we extracted from the line cut shown in Fig.~2(c). 
However, the fitted $\gamma_{rm} / 2\pi$ is larger than the expected value of 28~GHz/T. Possible reasons for the discrepancy include (1) that interactions between the spin wave modes and the upper (lower) branch distorts the shape of $\omega_{+}(B_0 > B_{res})$ and $\omega_{-}(B_0 < B_{res})$ from the model prediction of Eqn. \ref{omega} (see Fig 4(b)); (2) Although the fit requires a $\gamma_{rm}$, the resonances that we use to extract it are strongest for larger values of $\Delta$, where its influence is smallest.  Thus, we are not as sensitive to $\gamma_{rm}$ than we are to other parameters, and small covariances and distortions can lead to an unphysical value. Nevertheless, the most relevant parameters for this analysis, $g$ and $B_{res}$ are insensitive to the exact value of $\gamma_{rm}$.

\renewcommand{\thefigure}{S2}
\begin{figure}
\includegraphics{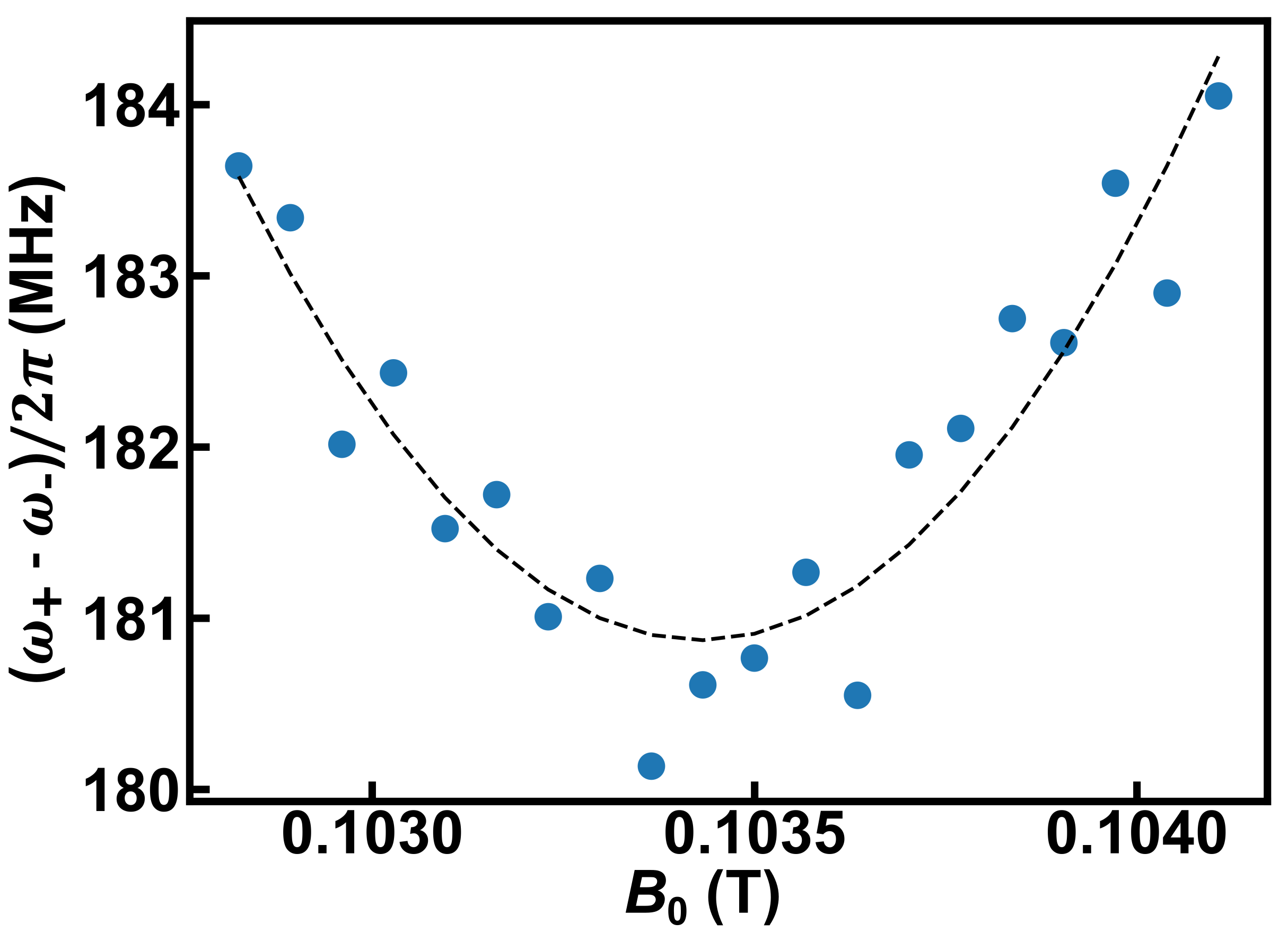}
\caption{Upper and lower branch frequency difference vs field and the fitting to extract $g$ and $B_{res}$. At the resonance field, the frequency difference is the smallest and is equal to $2g$.}
\end{figure}

\section{Magnon modes below the uniform magnon mode \\frequency}

In general, modes that appear at frequencies below the uncoupled uniform magnon mode frequency have been attributed to backwards volume modes \cite{Stenning:13,BhoiB2014Sopc}. Forward volume modes and surface spin waves have higher frequencies than the uniform magnon mode~\cite{serga2010yig}, so we do not consider them here as potential sources for these magnon modes below the uniform mode frequency.  Our experimental geometry makes the excitation and detection of quantized backwards volume modes unlikely because the magnetic field is applied parallel to the long axis of the magnetic material, with an estimated misalignment of less than 3$^\circ$. Because backwards volume modes require a wavevector parallel to the magnetic field, this would require quantization along the (600~$\mu$m) long axis and cannot explain the magnon modes below the uniform mode frequency because this scenario cannot impose visible quantization. We also note that the cross section of the sample is not rectangular at the edges due to the growth process~\cite{Franson2019}, which in some situations could give rise to a nonuniform demagnetization field, however this mechanism would also require a substantial magnetic field perpendicular to the long axis of the magnet, and is ruled out by our misalignment estimate above.  Finally, we have also considered the possibility of a nonlinear process in which a resonator photon couples to the uniform mode magnon and induces a transition from a uniform mode magnon into a 
$k\neq 0$ magnon.  This scenario is consistent with the appearance of modes at frequencies below the uncoupled uniform magnon mode. Our estimate of the resonator photon and magnon populations suggests that such a nonlinear process would be orders of magnitude smaller than observed. Due to these inconsistencies with respect to typical explanations for magnon modes below the uniform magnon mode, we cannot make a definitive assignment of the lower frequency quantized modes.

\section{Ring-down Experiments}

The setup for the ring-down experiments is shown in Fig S3. From the top-left of this diagram: the microwave (MW) signal generator generates a sinusoidal voltage at a frequency relevant to $\omega_+$ or $\omega_-$, which is then split using a directional coupler. The top branch, which we label the local oscillator, carries 11~dBm of MW power. The lower branch, which we label the signal carrier, carries $-$9~dBm of MW power. In the lower branch, a function generator outputs a square wave that controls a microwave switch that chops the lower branch microwave power in a square wave.  After the switch, the attenuation from the coaxial cable, a 20~dB attenuator outside the cryostat, and a 30 dB attenuator inside the cryostat attenuate the MW signal to $-$65~dBm before it reaches the sample. The signal from the sample is  amplified outside the cryostat before it's mixed with the local oscillator signal, with a power limiter before the mixer to protect it from overloading.  We adjust the phase shifter in the local oscillator branch so that the DC component of the mixed ring-down signal is the highest. At the end, a low pass filter rejects components at $2\omega$ and we are left with a DC component of the mixed signal that we digitize using a sampling oscilloscope. The resulting voltage is proportional to the amplitude of magnon-resonator ring-down, which we fit as discussed in the main text.

\renewcommand{\thefigure}{S3}
\begin{figure}
\includegraphics{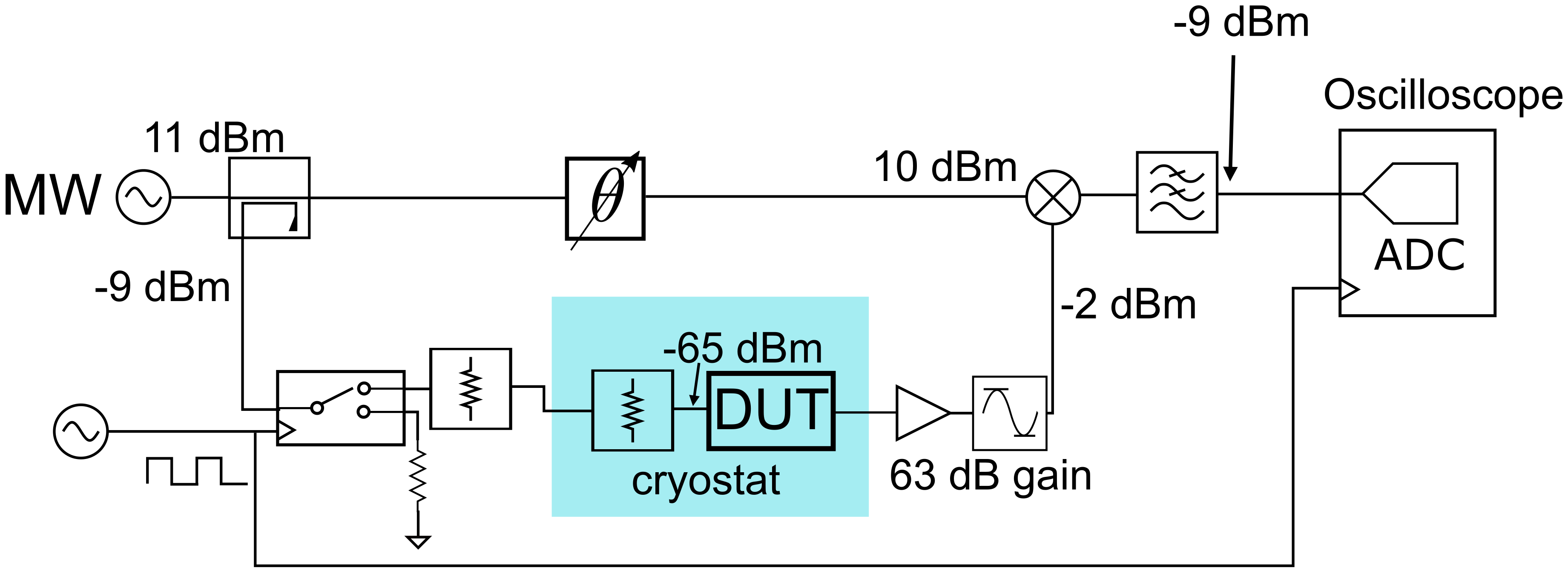}
\caption{Circuit diagram for the ring-down experiment.
}

\end{figure}

\section{Temperature Dependence}

Using the techniques discussed in the main text, we also investigate properties of our hybrid resonator-magnon system at higher temperatures. We plot $\kappa_m$ and $g$ in Fig.~S4(a) and S4(b) respectively. We see that $g$ decreases and $\kappa_m$ increases with increasing temperature.

\renewcommand{\thefigure}{S4}
\begin{figure}
\includegraphics{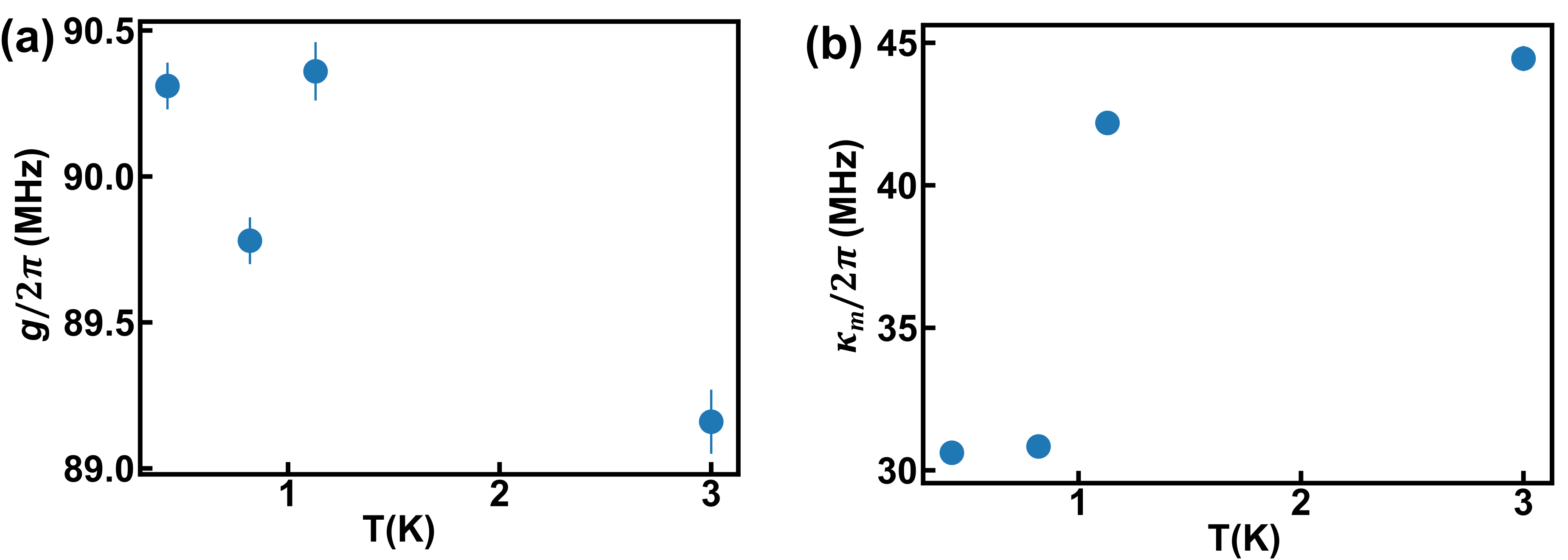}
\caption{MW measurement of \VTNCE~coupled to the 3.6~GHz resonator at different temperatures. (a) Temperature dependence of $g$ extracted from the $S_{21}$ measurements at $B_{res}$, (b) Temperature dependence of $\kappa_m$ extracted from $S_{21}$ linewidth measurements at $B_{res}$. The error bars for $\kappa_m$ are too small to be seen on this scale. 
}

\end{figure}

\renewcommand{\thefigure}{S5}
\begin{figure}
\includegraphics{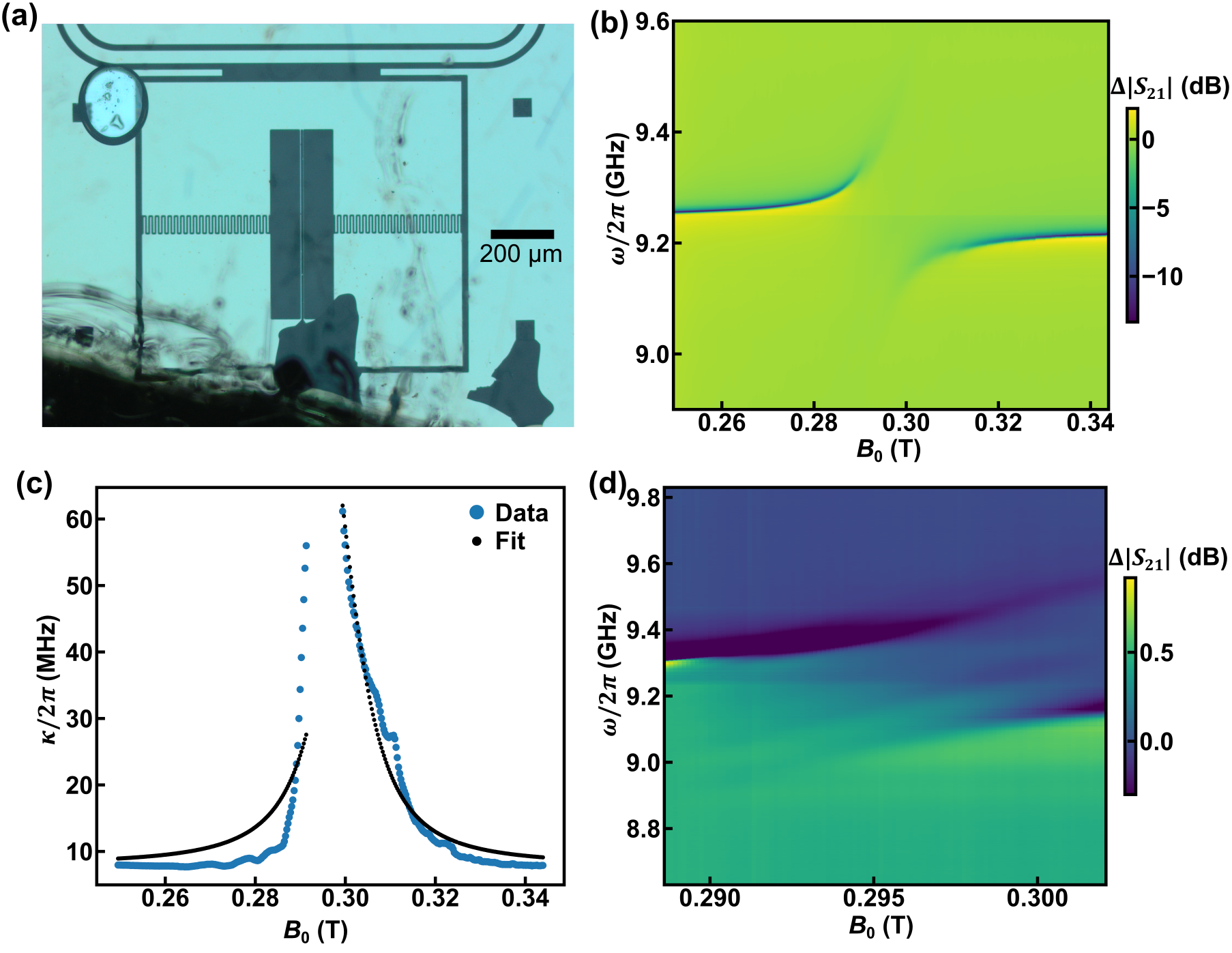}
\caption{MW measurement of \VTNCE~coupled to the 9.2~GHz resonator at 0.43~K. (a) Microscope image of \VTNCE~on the 9.2~GHz resonator. (b) $S_{21}$ measured as a function of $B_0$ and $\omega$. (c) Damping rate of the upper branch vs $B_0$ from 0.249~T to 0.291~T, and that of the lower branch from 0.299~T to 0.344~T. The fitted magnon damping rate is $\kappa_m = 117.7(5.3)$~MHz. (d) $S_{21}$ measurements with finer steps in $B_0$ showing spin wave modes around 150~MHz higher than the lower branch frequency.}

\end{figure}

\section{Second Device Results}

Other than the 3.6~GHz resonator as described in the main text, we also study another resonator-magnon device that is designed with a larger resonance frequency. It has the same inductor wire dimensions, but a smaller capacitance. The device resonance frequency is around 9.2~GHz, measured at zero field after V[TCNE]$_x$ deposition and encapsulation. A microscope image of the device after V[TCNE]$_x$ deposition is shown in fig S5(a). In this device there is incomplete lift-off.  A large area of \VTNCE~remains on the device, which is connected to the \VTNCE~strip on the inductor wire. This may result in the \VTNCE~strip coupled to a continuum of magnon modes.

Figure S5(b) shows $\Delta |S_{21}|$ as a function of magnetic field and frequency acquired at 0.43~K and $-$65~dBm of microwave power. As with the device discussed in the main text, we verify that we operate the device in the linear response regime; here between $-$55~dBm to $-$65~dBm.
 We again see a strong avoided level crossing, indicating that this device operates in the strong coupling regime. Both $\kappa_r$ and $\kappa_m$ are larger as compared to the 3.6~GHz device, and the upper and lower branch resonance signals are weaker at $B_0 = B_{res}$. We fit the data using Eqn. \ref{2dipsS21} of the main text to extract $\omega_+$, $\kappa_+$ from 0.249~T to 0.291~T, and $\omega_-$, $\kappa_-$ from 0.299~T to 0.344~T. As in the main text, we then substitute $\omega_\pm$ into Eqn. \ref{S21} of the main text and treat $g$, $\omega_{r0}$, $\gamma_r$ and $M_{\text{eff}}$ as free parameters. We obtain $g / 2\pi=147.21(29)$~MHz and $M_{\text{eff}}=72.40(11)$~mT, $\omega_{r0} / 2\pi = 9.2529(15)$~GHz and $\gamma_r / 2\pi = -71.7(5.1)$~MHz/T. Also, assuming linear dependence of $\kappa_r$ on $B_0$, we evaluate $\kappa_+(B_0 = 0.249$~T) $\approx \kappa_r(B_0 = 0.249$~T) and evaluate $\kappa_-(B_0 = 0.344$~T) $\approx \kappa_r(B_0 = 0.344$~T). We find $\kappa_r(B_{res}) / 2\pi= 7.917(15)$~MHz. 

To extract $\kappa_m$, we use~\cite{harder2021coherent}

\begin{equation}
    \tilde{\omega}_\pm =  \tilde{\omega}_r +  \tilde{\Delta}/2 \pm \sqrt{ \tilde{\Delta}^2 + 4g^2}/2,
\end{equation}
where $ \tilde{\omega} = \omega - i \kappa$/2 and $ \tilde{\Delta} =  \tilde{\omega}_m(B_0) -  \tilde{\omega}_r$. The real part of this equation gives Eqn. \ref{S21} when $\omega_{r,m} \gg \kappa_{r,m}$, and the imaginary part gives 
\begin{equation}
\kappa_\pm = [\kappa_r/2 + \kappa_m/2 \mp \text{Im} \sqrt{(-\omega_r+i\kappa_r/2+\omega_m- i\kappa_m/2)^2 + 4g^2}].
\end{equation}
 Using this expression of $\kappa_\pm$, and treating $\kappa_m$, $M_{\text{eff}}$ and $g$ as free parameters, we obtain $\kappa_m / 2\pi = 115$~MHz. The data and the fit are shown in Fig. S5(c). Therefore, this 9.2~GHz resonator-magnon hybrid system also operates in the strong coupling regime with $\mathcal{C} = 93.0(4.2)$. For the same system at 3.0~K, we extract $M_{\text{eff}} = 72.94(10)$~mT, $\kappa_r(B_{res}) / 2\pi = 10.456(31)$~MHz, $\kappa_m / 2\pi = 112.9(4.8)$~MHz, $g / 2\pi = 146.20(26)$~MHz. Therefore the cooperativity is $\mathcal{C} = 72.4(3.2)$.  

As in the main text, we measure $\Delta |S_{21}|$ with higher resolution around the avoided level crossing at 0.43~K [Fig. S5(d)]. We also see faint spin wave modes within the avoided-crossing gap between $\omega_{\pm}$.

\section{Magnetic Field Calibration}

To calibrate the magnetic field of the cryostat electromagnet at the sample location, we measure absorption electron spin resonance (ESR) of 2,2-diphenyl-1-picrylhydrazyl (DPPH) in contact with the central conductor of a broadband coplanar waveguide. DPPH is a stable radical with a well-studied gyromagnetic ratio across a broad range of temperatures. The ESR is measured at 4.7~K, where the $g$~factor is 2.083~\cite{voesch2015chip}. We measure ESR from 4.5~GHz to 8.5~GHz in steps of 0.5~GHz. At each frequency we sweep the magnet current as we acquire microwave transmission with a vector network analyzer.  We extract the magnet current associated with the spin resonance at each frequency. We find that the line formed by plotting the resonant currents as a function of frequency extrapolates to the origin as expected.  Using these data we calculate a calibration factor of 60.64(10)~mT/A with an uncertainty in absolute magnetic field of 0.4~mT. Fig. S6 shows a plot of the static field at the sample $B_0$ vs the magnet coil current.

\renewcommand{\thefigure}{S6}
\begin{figure}
\includegraphics{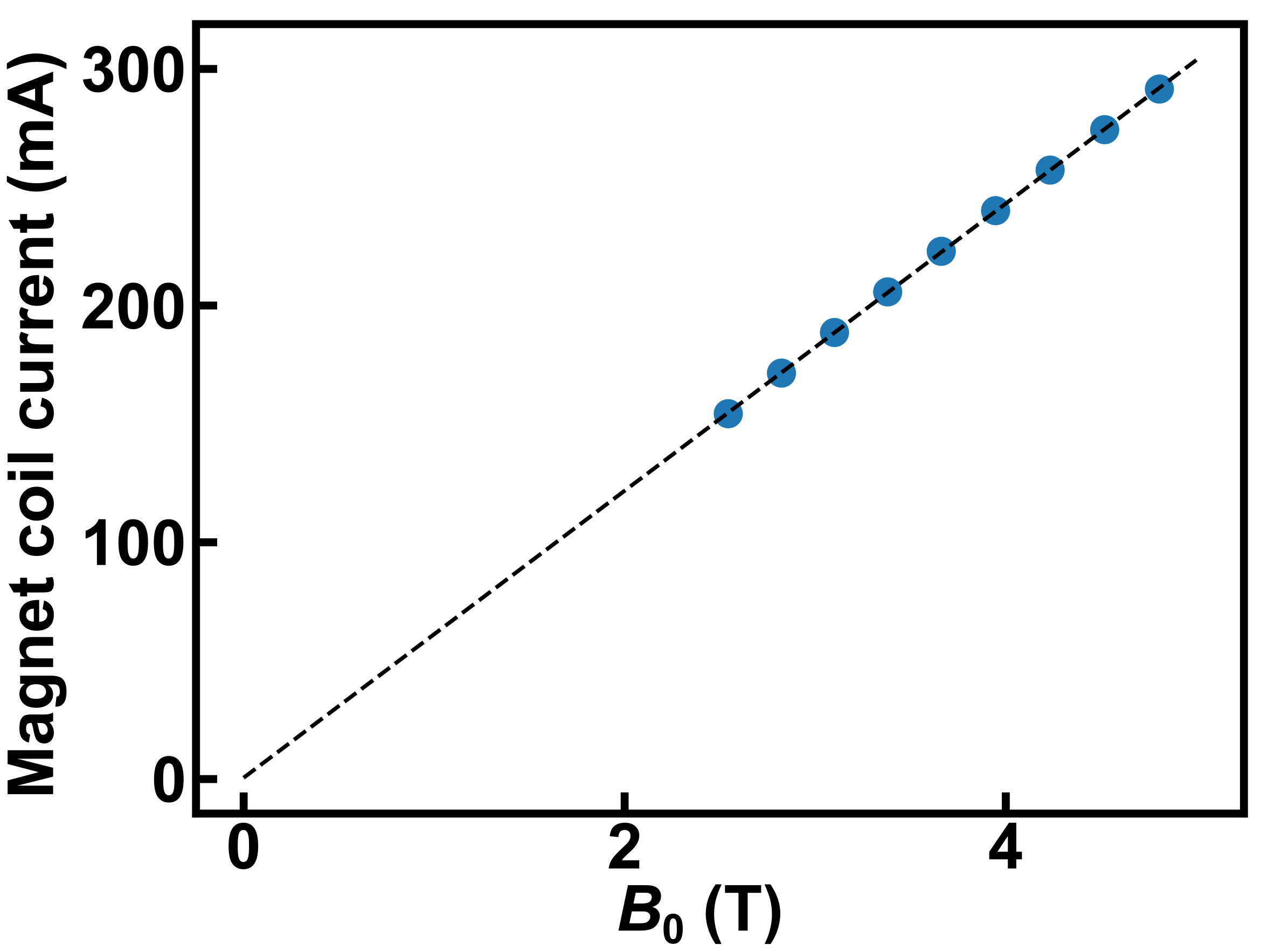}
\caption{Experimental data for the static field at the sample $B_0$ vs the magnet coil current (blue dots) and the linear fit (black dashed line) to extract the calibration factor of 60.64(10)~mT/A. The black dashed line extrapolates to the origin as expected.}

\end{figure}

\nocite{*}

\end{document}